\documentclass[dvips]{scrartcl}
\usepackage{amsmath}
\usepackage{slashed}
\usepackage{graphicx,epsfig}
\usepackage{hyperref}
\def\lsim{\mbox{\raisebox{-.6ex}{~$\stackrel{<}{\sim}$~}}}
\def\gsim{\mbox{\raisebox{-.6ex}{~$\stackrel{>}{\sim}$~}}}
\def\beq{\begin{equation}}
\def\eeq{\end{equation}}
\def\beqa{\begin{eqnarray}}
\def\eeqa{\end{eqnarray}}
\def\sfrac#1#2{{\textstyle{#1\over #2}}}
\def\L{{\scriptscriptstyle L}}
\def\R{{\scriptscriptstyle R}}

\title{Minimal hidden sector models\\ for CoGeNT/DAMA events}
\author{    James M.\ Cline$^1$ \thanks{jcline@physics.mcgill.ca},
 Andrew R.\ Frey$^2$ \thanks{a.frey@uwinnipeg.ca}
  \\
  \textit{\large $^1$ Physics Department, McGill University, Montreal, QC, H3A2T8,
    Canada}\\
\textit{\large $^2$Dept.\ of Physics and Winnipeg Insitute for Theoretical 
Physics,}\\ \textit{\large The
University of Winnipeg, Winnipeg, MB, R3B2E9, Canada}	
}
\date{}

\begin{document}
\maketitle

\begin{abstract} Motivated by recent attempts to reconcile hints of direct
dark matter detection by the CoGeNT and DAMA experiments, we construct 
simple particle physics models that can accommodate the constraints.  We
point out challenges for building reasonable models and identify the most
promising scenarios for getting isospin violation and inelasticity, as
indicated by some phenomenological studies.  If inelastic scattering is
demanded, we need two new light gauge bosons, one of which kinetically
mixes with the standard model hypercharge and has mass $< 2$ GeV, and
another which couples to baryon number and has mass $6.8\pm{\!_{0.1}\atop
\!^{0.2}}$ GeV.  Their interference gives the desired amount of isospin
violation.  The dark matter is nearly Dirac, but with small Majorana
masses induced by spontaneous symmetry breaking, so that the gauge boson
couplings become exactly off-diagonal in the mass basis, and the small
mass splitting needed for inelasticity is simultaneously produced. If only
elastic scattering is demanded, then an alternative model, with interference between
the kinetically mixed gauge boson and a hidden sector scalar Higgs, is adequate
to give the required isospin violation.  In both cases, the light
kinetically mixed gauge boson is in the range of interest for currently
running fixed target experiments.  

\end{abstract}

\section{Introduction}

Hints of direct detection of dark matter (DM) currently exist from two
experiments.  There is a long-standing observation of an annual
modulation in the signal observed by DAMA \cite{Bernabei:2008yi}, whose statistical
significance is beyond question.  Last year the CoGeNT experiment
reported excess events in their lowest electron energy bins
\cite{Aalseth:2010vx},
followed more recently by a $2.8\sigma$ detection of annual modulation
in the signal \cite{Aalseth:2011wp}.  Under the simplest assumptions about the nature of the
dark matter interactions, these two observations appear to be
incompatible with each other \cite{Savage:2010tg,McCabe:2011sr} and with upper limits obtained by other
experiments, especially CDMS \cite{Ahmed:2009zw,Ahmed:2010wy}, 
Xenon10 \cite{Angle:2009xb} and Xenon100 \cite{Aprile:2011hi}.  Channeling of recoiling
ions along the crystal planes in the detectors has been suggested as
one loophole for reconciling the conflicts, but this has been argued
to be too small an effect by ref.\ \cite{Bozorgnia:2010xy}. 
Uncertainties in quenching factors can also be used to help reconcile 
the two positive detections \cite{Hooper:2010uy}.

There are alternative microphysical ways of ameliorating the tensions that have
been explored in recent papers  
\cite{Chang:2010yk}-
\cite{Farina:2011pw}.  One important modification is to
allow for isospin-violating interactions of the dark matter with
nucleons, since the coherent spin-independent matrix element is a sum
over the interactions with protons and neutrons:
\beq
	\sigma_N \sim (f_p\, Z + f_n\, (A-Z))^2\, \sigma_n
\eeq
where $A,Z$ are the number of nucleons and protons in the nucleus,
respectively, and $\sigma_{n,N}$ are the respective cross sections 
for DM scattering on neutrons and the nucleus.  If isospin is
conserved then $f_p = f_n$, but if $f_n/f_p\sim -0.7$, the limits
placed on $\sigma_n$ by the Xenon experiments are greatly relaxed
\cite{Chang:2010yk,Feng:2011vu}.  

Moreover, several groups have indicated that inelastic scattering
\cite{TuckerSmith:2001hy} of DM with a small mass splitting $\delta
\sim 10$ keV  can have a beneficial effect either for the the
agreement between DAMA and CoGeNT \cite{Graham:2010ca}, the conflict
between CoGeNT and CDMS\footnote{Inelastic scattering increases the
ratio of the modulation amplitude to the unmodulated rate, softening
the discrepancy with CDMS \cite{Ahmed:2010wy}, which reports no
evidence of events that would be compatible with the CoGeNT signals.}
\cite{Chang:2010yk,Frandsen:2011ts}, the goodness of fit to
the CoGeNT modulated signal alone \cite{Fox:2011px}, or marginally
the overall goodness of fit to all data \cite{Farina:2011pw}. 

In this work, we look for the simplest and most natural kind of models
that could accommodate these two generalizations, isospin violation
and inelasticity, with an emphasis on hidden sector models with Higgs
or gauge kinetic portals to the standard model. (For previous
model building efforts which do not focus on these aspects, see references
\cite{Fitzpatrick:2010em}-\cite{Buckley:2010ve}.)  A similar study to the present one was
done in ref.\ \cite{Gao:2011ka}, but considering more elaborate models
than we examine here.  
Our inelastic model also has elements in common with
that of ref.\ \cite{Gondolo:2011eq}, although the latter incorporated
neither isospin violation nor inelasticity.  (See \cite{DelNobile:2011je} for
another recent isospin-violating model.)  In ref.\ \cite{Frandsen:2011cg}, isospin 
violating couplings of a single  vector were obtained through a
combination of kinetic mixing and mass mixing with the $Z'$.
 We will argue that
interference between two new vector mediators is an elegant way
of getting isospin violation if one demands that only inelastic scattering
takes place. It is well known that Dirac states
coupling to vectors become off-diagonal in the mass eigenstate basis
if small Majorana masses are introduced \cite{TuckerSmith:2001hy}.  
Thus it is not challenging to account for the inelastic nature of the
couplings.

On the other hand, ref.\ \cite{Farina:2011pw} finds that inelasticity
improves the global fit to the data only moderately, so that one might
also contemplate models with no DM mass splitting and purely elastic
scattering.  The overall fit is in fact not very good, indicating
either that some of the data have inconsistencies or that the dark
matter interpretation is not correct.  In this work we will assume the
former, in which case one might be motivated to consider inelasticity
as a secondary criterion, which might or might not survive as the data
improve.  Accordingly, we consider both elastic and inelastic models
in the following and leave it to the reader to judge how strongly the
latter is preferred by the data. Elastic scattering naturally arises in
alternative models where the isospin violation comes about by
interference between vector and scalar  exchange.  We will construct
our models for isospin-violating dark matter starting with this
simpler possibility, showing why it does not naturally accommodate
inelastic couplings.

\section{Scalar versus vector exchange}

A natural way to induce scalar-mediated
interactions between the dark sector and the standard model (SM) is
to introduce a new Higgs field $\phi$ that is a singlet under
the SM gauge group, and communicates to the SM through the
interaction $\lambda \phi^2 h^2$.  If $\phi$ gets a VEV then
the mass eigenstates are admixtures of $\phi$ and $h$ with mixing
angle $\theta_s \cong \lambda \langle \phi\rangle v(m^{-2}_h -
m^{-2}_\phi)$, where $m_{h,\phi}$ stand for
the mass eigenvalues (we assume small mixing, $\theta_s\ll 1$)
and $v=246$ is the SM Higgs VEV.  If the DM $\chi$ has a Yukawa
coupling $y\bar\chi\phi\chi$, then $\phi$ mediates interactions
with SM fermions $f$ with the strength $y y_f \theta_s / m^2_\phi$,
assuming that $m_\phi\ll m_h$.  Here $y_f$ is the SM Yukawa coupling
of the Higgs to fermion $f$.  

In this scenario, $\phi$ couples to SM matter proportionally to
$h$.  Although these couplings violate isospin, it is a very small
violation when it comes to scattering on nuclei, because nucleons get
almost none of their mass from the valence quarks.  The couplings of
the Higgs to nucleons are dominated by the sea-quark and gluon
content, which are the same for neutrons and protons (see for example
\cite{Barger:2010yn,Gao:2011ka}).  Therefore we
get no significant isospin violation from couplings mediated by the
Higgs portal: it has $f_p = f_n$.

If in addition we invoke a gauge kinetic mixing portal,
$(\epsilon/\cos\theta_W)
F_{\mu\nu} Z'^{\mu\nu}$ (where $F$ and $Z'$ are the respective field strengths of
the SM hypercharge and the hidden sector U(1)$'$, and
$\theta_W$ is the Weinberg angle),
then interference between the new vector and the new Higgs does lead
to tunable isospin violation.  This is because the vector only mixes
significantly with the photon, leading to $f_n/f_p=0$.
Of course, this is not the ratio we need for DAMA and CoGeNT.  But
interference between the vector and scalar exchange allows for any 
desired value.  If $g'$ is the new U(1)$'$ gauge coupling to the DM, then
\beq
	{f_n\over f_p} \cong {(y y_n \theta_s/m_\phi^2)\over 
	(g' e \epsilon/m_{Z'}^2) + (y y_n\theta_s /m_\phi^2)}
\label{frat}
\eeq
where $y_n$ is the coupling of $h$ to the nucleon:
$y_n\cong 0.36\, m_n/v$ in terms of the nucleon mass $m_n$ and the
Higgs VEV $v$ \cite{Ellis:2000ds,Giedt:2009mr}.

The main objection to this scenario arises if we want the couplings to
DM to also be inelastic. Suppose there are two mass eigenstates
$\chi_\pm$ with masses $M_\pm = M\pm\delta/2$ split by the small amount
$\delta$.  Let us first write down an effective
potential:
\beq
	V = \sum_\pm \bar\chi_\pm M_\pm\chi_\pm 
	+\frac12 m^2_{Z'} {Z'}^2 + \frac12 m^2_\phi \phi^2
	+g\bar\chi_+ \slashed Z' \chi_-
	+ y \bar\chi_+ \phi \chi_-
\label{scalarV}
\eeq
(It is also understood that $Z'$ couples to the electromagnetic current
with strength $\epsilon e$ and $\phi$ to the SM fermions $f$ with 
strength $\theta_s y_f$.)  The mass splitting can arise by spontaneous
symmetry breaking through couplings of the form 
$\phi^*(y_\L\chi_\L\chi_\L + y_\R \chi_\R\chi_\R)$
where $\chi_{\L,\R}$ are the Weyl components in the original Lagrangian,
and $\phi$ carries twice the U(1)$'$ charge of $\chi$.

The troublesome question is how the scalar
interactions came to be purely off-diagonal in the mass eigenbasis,
given that $\phi$ has a VEV which is needed in order to mix with the SM
Higgs $h$.
To arrive at (\ref{scalarV}), there must have been a bare
mass term $\mu\bar\chi_+\chi_-$ (here expressed in the mass eigenbasis) that was exactly canceled by 
$y\langle\phi\rangle \bar\chi_+\chi_-$.  In the absence of this tuning,
$\phi$ will have diagonal plus off-diagonal couplings, and this presents a
complication for getting the desired level of isospin violation, since the
interference between $\phi$ and $Z'$ only occurs in the inelastic
channel.  This is because the Majorana vector couplings are purely
off-diagonal (see the following subsection). We do not
contemplate this finely-tuned situation any further.   

However if instead of couplings of the form $\phi^*(\chi_\L\chi_\L +
\chi_\R\chi_\R)$ in which $\phi$
must carry a compensating charge, we have the interactions
\beq
 \bar\chi\left(M+y\phi\right)\chi
\eeq 
in terms of a Dirac fermion $\chi$,
then $\chi$ remains Dirac after $\phi$ gets its VEV, and there
is no mass splitting.  $\phi$ can interfere with $Z'$ through
the purely elastic couplings.  Even though we must invoke an additional
singlet Higgs $\tilde\phi$ to break the U(1)$'$ symmetry, since now $\phi$
is neutral under U(1)$'$, this is still an economical model, whose
consequences we will consider.  For simplicity we take $\phi$
to be real.  We do not consider the case of a pseudoscalar 
($i\gamma_5$) coupling because this leads to a nuclear scattering 
amplitude that is suppressed
by the DM velocity, making it more difficult to interfere with the
vector exchange contribution to get the desired isospin violation 
(since the latter has no such velocity 
suppression).

\subsection{Nondiagonal gauge couplings}

In contrast to the couplings of a scalar to DM, 
the purely off-diagonal coupling of the gauge field
is natural if $\chi_\pm$ are Majorana fermions that originated from
a Dirac particle before spontaneous symmetry breaking
\cite{TuckerSmith:2001hy}.  Consider
the interactions (again in the model where $\phi$ carries two units of 
the $\chi$ charge)
\beq
	V = \frac12 \bar\chi_\L M \chi_\R + \frac{y}{2}\phi^*\,
	\left(  
	\bar\chi_\L\,  P_L\, \chi_\L +  \bar\chi_\R\, 
	P_R\, \chi_\R\right) +{\rm h.c.}
\label{lagrangian}
\eeq
where now $\chi_L^T = (\psi_\L, \sigma_2 \psi_\L^*)$, 
$\chi_\R^T = (-\sigma_2\psi_\R^*, \psi_\R)$  denote Majorana-Dirac spinors
constructed from the Weyl spinors, here renamed $\psi_{\L,\R}$ to avoid
confusion.  When $\phi$ gets a
VEV, the mass matrix becomes 
\beq
	\left(\begin{array}{cc}\mu & M\\ M& \mu\end{array}\right)
\eeq
which is diagonalized by $\chi_{\L,\R} = \frac{1}{\sqrt{2}}(\chi_+\pm
\chi_-)$ with mass eigenvalues $|M_\pm| =  M \pm \mu$, where $\mu = 
y\langle\phi\rangle \ll M$.  If the U(1)$'$ interaction was orginally
vector-like, then it becomes exactly off-diagonal in the mass basis
because there is no vector current for a single Majorana state.
This is a strong motivation to prefer vector mediators if we aim for 
both isospin violation and inelastic off-diagonal couplings.

\section{Interfering vector exchanges}

The previous discussion motivates us to build a model in which the 
interfering
scalar current is replaced by another vector current.  The new vector need only
couple to isospin differently from the kinetically mixed U(1) that
has $f_n/f_p=0$.  Coupling to $B-L$ is attractive from the point of
view of anomaly cancellation, but such couplings are very strongly 
constrained because of the leptonic interactions (see for example
\cite{Williams:2011qb}).  The simplest
possibility that avoids these constraints is coupling to $B$ alone.
U(1)$_B$ is anomalous and it also has mixed anomalies with the SM
gauge groups, that can be canceled by adding the appropriate exotic
heavy particles \cite{Carone:1994aa}-
\cite{Pospelov:2011ha}.  We will 
not discuss the implications of these new particles further here,
although they can provide complementary collider signatures to test
the model.  Our addition of a single vector-like DM particle coupling
to $B$ does not spoil the anomaly cancellation achieved in these models.

We refer to the U(1)$_B$ gauge boson as $B_\mu$ and for simplicity
assume that it couples with equal strength $g_B$ to the DM and to the
SM baryons.  It also couples with equal strength to protons
and neutrons, just like the singlet Higgs of the previous section.
Therefore it is clear that (\ref{frat}) is replaced by
\beq
	{f_n\over f_p} \cong {(g_B^2 /m_{B}^2)\over 
	(g' e \epsilon/m_{Z'}^2) + (g_B^2 /m_{B}^2)}
\label{frat2}
\eeq

This model is almost complete, but we have accounted for the breaking
of only one linear combination of the two new U(1)s through the VEV of
$\phi$.  Notice that $\phi$ must have charges $-2(g',g_B)$ under
$U(1)'\times U(1)_B$ in order for (\ref{lagrangian}) to be gauge invariant. 
To completely break the symmetry we need another field $\tilde\phi$
with different charges.  This means that the mass eigenstates for the
gauge bosons are generally admixtures of the original fields, and that
both will therefore kinetically mix with the SM hypercharge.  We need
to clarify the relation between the couplings appearing in (\ref{frat2})
and the original Lagrangian parameters.

A simple way to ensure that the above relations are approximately correct
despite mixing of the new gauge bosons is to assume that $m^2_B \gg
m^2_{Z'}$,
by assigning $\tilde\phi$ the charges $(0,\tilde g_B)$
such that $\tilde g_B^2\langle\tilde\phi\rangle^2 \gg 
g_B^2\langle\phi\rangle^2$.
In that case the mixing is suppressed by
the large mass difference\footnote{If $\langle\phi\rangle=u$ and 
$\langle\tilde\phi\rangle = \tilde u$, then the gauge boson mass matrix
in the basis $(B,Z')$ is 
$\left({\tilde g_B^2\tilde u^2 + 4g_B^2 u^2 \atop 4g_B g'u^2}\,
{4g_B g'u^2 \atop 4g'^2 u^2}\right)$} and we can take 
\beq
	m_B \cong \tilde g_B \langle\tilde\phi\rangle \quad \gg \quad
	m_{Z'} \cong 2\, g' \langle\phi\rangle
\label{masses}
\eeq
The mixing angle is approximately $\theta \cong
(g_B/g')(m_{Z'}/m_B)^2$.  We will find that its value scales
proportionally to the gauge kinetic mixing parameter $\epsilon$, such that
$\theta\cong 4\epsilon$.  (This relation follows from eq.\ (\ref{mbzconst})
below and the relic density constraint, fig.\ \ref{fig:relicden}(a).)

\begin{table}[t]
\begin{center}
\begin{tabular}{|c|c|c|c|c|}
\hline
Model & $\sigma_n$ (cm$^2$)$^{\phantom|}$ & $M$ (GeV) & $\delta$ (keV) & $f_p/f_n$ \\
\hline
vector $B_\mu$ exchange & $3\times {10^{-38}}^{\phantom|}$ & 8 & 9.3 & $-1.53$ \\
\hline
scalar $\phi$ exchange & $6\times {10^{-39}}^{\phantom|}$ & 7.5 & 0 & $-1.54$ \\
\hline
\end{tabular}
\end{center}
\caption{\label{tab1} Best-fit values of the 
DM-neutron elastic scattering cross section, DM mass, mass splitting, and
isospin violation, from ref.\ \cite{{Farina:2011pw}}, appropriate to the
given theoretical model.}
\end{table}

\section{Determining the couplings} 

We must show that values for the parameters exist that can give the
right cross section for CoGeNT and DAMA, and the right relic density
for the dark matter.\footnote{Since our DM is Dirac or quasi-Dirac, there
is the interesting possibility for an asymmetry between $\chi$ and $\bar\chi$ being
the origin of the relic density, which has been widely discussed in 
the recent literature (see for example \cite{Kaplan:2009ag}).  
For this work we will assume the asymmetry vanishes.}
 For the effective elastic cross section of
DM on the neutron, the DM mass and mass splitting, and level of
isospin violation, we will consider the three cases shown in table
\ref{tab1},
which is corresponds to the best-fit values found by ref.\ 
\cite{Farina:2011pw} for the cases of endothermic $\chi_-N\to \chi_+N$,
and elastic scatterings, respectively.  These
are the ones appropriate to our two models.    
We will show that the exothermic reactions, $\chi_+ N\to
\chi_-N$, are  possible when $\epsilon \gsim 10^{-3.5}$,
whereas otherwise the excited state is 
depleted by $\chi_+\chi_+\to\chi_-\chi_-$ downscatterings in the
early universe.

The theoretical cross sections for DM-neutron scattering in our models, in the elastic limit, are
\beq
	\sigma_n = {\mu_n^2\over \pi}\times\left\{\begin{array}{rl}
	{g_B^4/m_B^4},  & B_\mu{\rm\ exchange}\\
	(y y_n\theta_s)^2/m_\phi^4,& {\rm\ scalar\ }\phi {\rm\ exchange}\\
\end{array}
	\right\}
\label{sigma_n}
\eeq
where $\mu_n$ is the reduced mass for the DM-nucleon system and $\vec v$
is the DM velocity.
By equating the $\sigma_n$ values  in table \ref{tab1} to those in 
eq.\ (\ref{sigma_n}) 
and using eqs.\ (\ref{frat},\ref{frat2}), we obtain 
\beqa
	{m_B\over g_B} &=& 232 {\rm\ GeV},\quad
	{m_{Z'}^2\over g'\epsilon} = -(79.9{\rm\ GeV})^2\quad (B_\mu{\rm\
exchange})\label{mbzconst}\\
	{m_\phi\over\sqrt{y y_n \theta_s}} &=& 346 {\rm\ GeV},\quad
	{m_{Z'}^2\over g'\epsilon} =  -(118.9{\rm\ GeV})^2\quad 
	({\rm scalar\ }\phi{\rm\ exchange})\label{mphizconst}
\eeqa
To get the correct mass spectrum for the model with $B_\mu$, we can set the bare Dirac mass
directly to $M=8$ GeV, and choose $y\langle\phi\rangle = 4.7$ keV.
Recall that only $Z'$ gets its mass primarily from $\phi$, in this model,
so if $\epsilon$ is sufficiently small, it is possible to have
$\langle\phi\rangle\sim 10$ GeV or less.  
The Yukawa coupling still needs to be quite small in that case, $y \sim
0.5\times 10^{-6}$.  However this is only 4 times smaller than the electron
Yukawa coupling in the standard model, so it is not unreasonable. 
Alternatively, for the purely elastic models with $\phi$ exchange, 
we need to set $M + y\langle\phi\rangle =$ 7.5 GeV.

\subsection{Relic density}

Next we consider the relic density. Starting with the $B_\mu$ vector exchange
model, there are several possible 
annihilations into gauge bosons:
$\chi_\pm\chi_\pm\to BB$, $\chi_\pm\chi_\pm\to Z'Z'$, 
$\chi_\pm\chi_\pm\to BZ'$.  In addition there are coannihilations
into quarks, $\chi_+\chi_-\to q\bar q$ mediated by $B$ in the
$s$-channel, and also  $\chi_+\chi_-\to f\bar f$ into all charged
SM fermions $f$ except the kinematically inaccessible top, mediated by the $Z'$.  Averaging over all the
possibilities, we find the annihilation cross section 
\beq
	\langle \sigma_{\rm ann} v\rangle = {1\over 32\pi\, M^2}\,
	S(g',g_B,\epsilon,x',x_B)
\label{sigv}
\eeq
where $S$ is a dimensionless function of the couplings and the mass
ratios $x' = m_{Z'}/M$, $x_B = m_B/M$, given by
\beqa
	S &=& \sfrac12\, g'^4\, f_1(x',x') + g_B^2\, g'^2\, f_2(x',x_B) + 
	\sfrac12\, g_B^4\, f_1(x_B,x_B) \nonumber\\
	&+&  g_B^4\!\!\!\! \sum_{i=u,d,s,c,b}\!\!\! N_{c,i}\, f_3(x_B,x_i) 
	\ +\ 
	(g'\epsilon e)^2\!\!\!\! \sum_{i={e,\mu,\tau,\atop u,d,s,c,b}}
	\!\!\! N_{c,i}\, Q_i^2\, f_3(x',x_i)
\label{Xeq}
\eeqa
Here $x_i = m_i/M$ for SM fermion $i$, with charge $Q_i$ and
number of colors $N_{c,i}$, and the functions $f_i$ are given in
the appendix.

For the the scalar exchange model, all the analogous processes to the previous
case are present, with $B_\mu$ replaced by $\phi$.  The $\chi\chi\to\phi
\phi$ contribution is $p$-wave suppressed, as is $\chi\chi\to f\bar
f$ by $\phi$ exchange, so we neglect them.  The
$\chi\chi\to Z'Z'$ and $Z'$-mediated $\chi\chi\to\bar f f$ contributions
are the same as in (\ref{Xeq}).  We find that (\ref{Xeq}) is replaced by
\beq
	S \to \sfrac12\, g'^4\, f_1(x',x') + y^2\, g'^2\, f_4(x',x_\phi)
	+ 
	(g'\epsilon e)^2\!\!\!\! \sum_{i={e,\mu,\tau,\atop u,d,s,c,b}}
	\!\!\! N_{c,i}\, Q_i^2\, f_3(x',x_i)
\label{Xeq2}
\eeq
where $x_\phi = m_\phi/M$ and $f_4$ is defined in the appendix.

\begin{figure}[t]
\centering
\includegraphics[width=\textwidth]{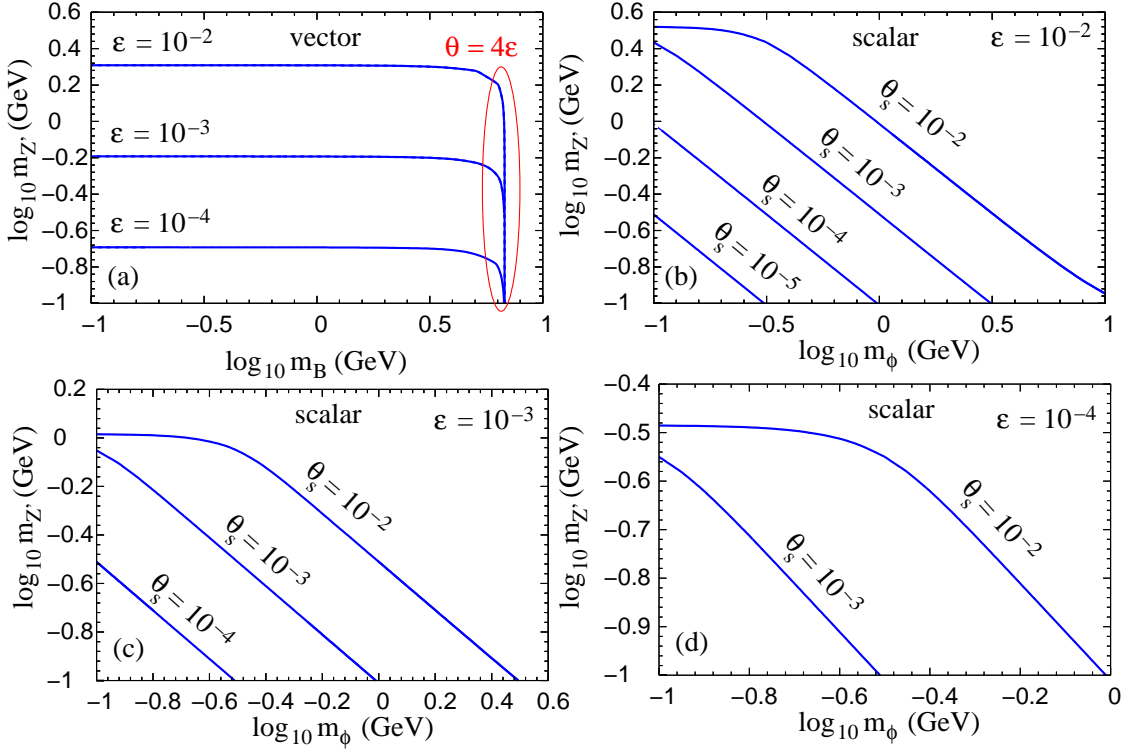}
\caption{\label{fig:relicden} (a): 
Contours in the  $m_{B}$-$m_{Z'}$ plane that give the observed relic
density in the vector exchange model,  for gauge kinetic mixing parameter 
$\epsilon = 10^{-2}$, $10^{-3}$, and $10^{-4}$. The ellipse highlights the
region where the gauge boson mixing angle $\theta$ is small as needed for
consistency.\newline
 (b)-(d):  Analogous contours in the
$m_{\phi}$-$m_{Z'}$ plane for the scalar exchange model, for 
$\epsilon = 10^{-2}$ (b), $10^{-3}$ (c), $10^{-4}$ (d), and 
several values of the Higgs mixing angle $\theta_s$.}
\end{figure}

For each of the models, we equate (\ref{sigv}) to the 
standard value of the cross section for 
the observed relic density, $\langle \sigma_{\rm ann} v\rangle
= 3\times 10^{-26}$ cm$^3$/s.  This gives the constraint
\beq
	S = 1.7\times 10^{-5}
\label{Xconst}
\eeq
To determine the parameters satisfying (\ref{Xconst}) in the vector
exchange model, we assume several
choices for $\epsilon$ that can be compatible with laboratory  bounds
on kinetic mixing (see next section), and use (\ref{mbzconst}) to
eliminate $g',g_B$ in favor of $m_{Z'},m_B$.  For the scalar exchange
model, we similarly use (\ref{mphizconst}) to eliminate $y$  in favor of
$m_\phi$.  This case has the free parameter of the Higgs
mixing angle $\theta_s$ to be varied in addition to $\epsilon$.
For $m_\phi<$ 5 GeV $\theta_s$ is constrained to be less than $0.01$
{}from the width of the $Z$ boson due to decays $Z\to \phi f\bar f$
and from $B$ meson decays $B\to \phi f\bar f$.
\cite{O'Connell:2006wi}.

We thus obtain, for
each value of $\epsilon$, the contour in the $m_{Z'}$-$m_B$ plane
corresponding to the observed relic density for the vector exchange model,
shown in fig.\ \ref{fig:relicden}(a).  
Similarly in the scalar  model,  
for each pair $\{\epsilon,\,\theta_s\}$, we find a contour in the
$m_{Z'}$-$m_\phi$ plane, shown in fig.\ \ref{fig:relicden}(b)-(d).
It is clear from fig.\ \ref{fig:relicden} that $m_{Z'}$ and $m_{\phi}$
tend to be $< 1$ GeV and 
both  scale as $\sqrt{\epsilon}$.   Only the vector $B_\mu$
can remain somewhat heavier as we now explain.

Recall that we previously made a simplifying technical assumption,
$m_B\gg m_{Z'}$, to ensure small mixing $\theta$ between the two gauge
bosons. This assumption might be relaxed somewhat, but at the risk
of increasing the  highly constrained couplings of $B$ to leptons due to the gauge kinetic
mixing.  To the extent that $\theta$ is small, only the vertical
part of the contours where $m_B = 6.8$ GeV is relevant, giving a 
sharp prediction for the $B_\mu$ mass, if $M$ is known.  Since
the determination of $M$ could well be uncertain by $\pm$1 GeV, we find
an uncertainty of $\pm{\!_{0.1}\atop \!^{0.2}}$ GeV in $m_B$ by varying 
$M$.  In this
vertical branch, the $\chi\chi$ annihilation cross section is dominated 
by the $g_B^4$ contributions in (\ref{Xeq}). 

\subsection{Relative abundance of excited state}

In the vector exchange model where we have a small DM mass splitting,
the process $\chi_+\chi_+\to \chi_-\chi_-$ mediated by the $Z'$
(and the $B$, although we find the former dominates) 
efficiently depletes the $\chi_+$ population in the early universe
over part of the allowed parameter space.
The downscattering cross section for a similar model was calculated in 
\cite{Cline:2010kv}, which adapts to the present case as
\beq
	\langle\sigma_\downarrow v\rangle \cong 
	\left({g'^2\over m_{Z'}^2+2M\delta} + {g_B^2\over m_{B}^2 +
2M\delta}\right)^2
	\,  {M^2\over 
	4\pi}\, \sqrt{2\delta\over M}
\eeq
These interactions freeze out at temperature $T_f$ given by 
$n\langle\sigma_\downarrow v\rangle = H$ where $n\sim (7\times
10^{-10}{\rm GeV}/M) T_f^3$ is the DM
number density and $H\sim T_f^2/M_p$ is the Hubble constant. 
If the DM remained in kinetic equilibrium with the SM down to
$T_f$ then the relative abundance of $\chi_+$ to $\chi_-$ would
be suppressed by $\sim \exp(-\delta/T_f)$.  However it is the
kinetic temperature of the DM which is important here, and if
kinetic decoupling occurs at a temperature $T_d > T_f$, then the
suppression is more severe, $\sim \exp(-\delta T_d/T_f^2)$
(see \cite{Cline:2010kv} for a discussion of this issue.)  The
kinetic equilibrium is controlled by the scattering of DM on
electrons, whose cross section is approximately
\beq
	\langle\sigma_{\chi e} v\rangle \cong 
	\left({g'\epsilon e\over m_{Z'}^2+2M\delta}\right)^2
	\,  {m_e^2\over 
	\pi}\, \sqrt{2T\over M}
\eeq
We find that this goes out of equilibrium at $T_d\cong 10^3\delta
\cong 10$ MeV.  

Using this methodology, we estimate that the relative abundance of
$\chi_+$ is unsuppressed for $\epsilon \gsim 10^{-3.5}$. The
suppression turns on exponentially fast as a function of $\epsilon$,
with $\epsilon=10^{-4}$ giving a relative abundance of
$\sim\exp(-10^4)$, while $\epsilon=10^{-3}$ leads to almost no
suppression, $e^{-0.1}$.    Therefore it is possible to realize the
exothermic dark matter scenario suggested in ref.\ 
\cite{Graham:2010ca} over some part of the allowed parameter space.
In these cases the scatterings will be an average over the endothermic
and exothermic ones since both states are equally populated.  We note
that ref.\ \cite{Farina:2011pw} finds a moderate preference for
exothermic reactions in their global fits.

\section{Discussion}

To recapitulate, we have investigated two hidden sector dark matter models
that violate isospin in an optimal manner for reconciling the CoGeNT and
DAMA signals with constraints from Xenon10 and Xenon100.  
The vector exchange model has two Majorana mass eigenstates 
$\chi_\pm$ (or a single pseudo-Dirac particle) 
with masses $M_\pm = M\pm\delta\cong 8$ GeV $\pm 5$ keV, 
and two light gauge bosons $Z'$ and $B$ with masses $m_{Z'} < 2$ GeV,
$m_B \cong 6.8$ GeV, and the couplings\footnote{There are also couplings
to the $Z$ boson current $j^\mu_{Z}$ given by 
${-\epsilon\tan\theta_W\, m_Z^{-2}}\, j^\mu_{Z}\left(m^2_{Z'}\,Z'_\mu 
+ \theta \, m^2_B\, B_\mu\right)$ 
 that come from the mixing of $Z'$ with weak hypercharge;
see for example \cite{Chen:2009ab}.  Larger contributions to the
kinetic mixing of the $B$ with $Z$ can be generated from SM
loops below the scale of baryon symmetry breaking \cite{Carone:1995pu}.}
\beq
	\bar\chi_+(g'\slashed{Z'} + g_B\slashed{B})\chi_- + g_B j_B^\mu B^\mu
+ \epsilon\, j^\mu_{EM}(Z'_\mu + \theta B_\mu) 
\eeq
Here $\epsilon<10^{-2}$ is the gauge kinetic mixing parameter for 
the $Z'$, $\theta \cong 4\epsilon$ is the mixing angle between $B$
and $Z'$, and $j^\mu_{EM,B}$ are the respective electromagnetic
and baryon number currents of the standard model.
The couplings
are adjusted to give the optimal isospin violation $f_p/f_n\cong -1.5$
\cite{Farina:2011pw} via eq.\ (\ref{frat2}).  There must also be diagonal Yukawa 
interactions $y\bar\chi_\pm\phi\chi_\pm$ to a singlet Higgs $\phi$
whose VEV leads to the small mass splitting, but $y\sim 10^{-5}$ is much 
smaller than the gauge couplings ($g_B=0.029$) and therefore we have neglected these
interactions.  The Dirac mass $M=8$ GeV appearing in the original
Lagrangian  is protected by chiral symmetry, and so does not
introduce any new hierarchy problem.  The small scale $\langle
\phi\rangle \sim$ GeV on the other hand is unexplained unless
one invokes supersymmetry in the hidden sector \cite{Cheung:2009qd}
or some other UV completion.

The scalar  exchange model has exactly 
Dirac dark matter with $M=7.5$ GeV  (hence only elastic scattering)
and a real singlet
$\phi$ that mixes with
the SM through the Higgs portal $\lambda \phi^2 h^2$.  Like the
previous model these also have the kinetically mixed $Z'$ vector.  The
interactions are given by
\beq
	\bar\chi(g'\slashed{Z'} + y\phi)\chi + \theta_s \sum_i y_i
	\bar f_i \phi f_i + \epsilon\, j^\mu_{EM}Z'_\mu 
\eeq
where $f_i$ are the SM fermions with their Yukawa couplings $y_i$.
We find that $m_\phi \lsim 10$ GeV to satisfy the relic density
constraint.
 If $m_\phi$ happens to be close to this upper limit, it could
be discoverable at the LHC, while for the very light cases $m_\phi<$ 1 GeV,
indirect discovery could come from rare decays such as
$B\to \phi X$ followed by $\phi\to \mu^+\mu^-$ \cite{O'Connell:2006wi}.

One of the most exciting aspects of these models is that they predict
new low-energy interactions mediated by the light $Z'$ with a strength
relevant for detection in beam-dump experiments 
\cite{Bjorken:2009mm,Essig:2009nc} such as APEX \cite{Essig:2010xa}
and the Mainz Microtron \cite{Merkel:2011ze} and the low-energy
$e^+$-$e^-$ collider experiment KLOE \cite{Archilli:2011nh}.
The still-open window of parameter space in the $\epsilon$-$m_{Z'}$
plane corresponds roughly to that which we have identified in this
paper as being compatible with the $\chi_\pm$ relic density.
In our model, $m_{Z'}$ is only bounded from above, depending on the value
of $\epsilon$, as shown in fig.\ \ref{fig:relicden}: $m_{Z'}\lsim
2\sqrt{\epsilon/10^{-2}}$ GeV in the vector exchange model.

On the other hand, the gauge boson $B$ of baryon number is extremely
hard to detect due to its very weak coupling and relatively large mass.
For example, constraints from new contributions to $\Upsilon$ decay into
quarks are easily satisfied \cite{Aranda:1998fr}.  The Tevatron sets
limits on $g_B\lsim 0.6$ from the nonobservation of $p\bar p\to B_\mu^*\to
\chi \chi j$ where $j$ is a single jet \cite{Graesser:2011vj}, which 
is also satisfied by our model.  Ref.\ \cite{Carone:1995pu}
pointed out that the kinetic mixing of $B$ leads to weak Tevatron
constraints from the Drell-Yan production of lepton pairs.  It may be
interesting to update these constraints since \cite{Carone:1995pu} was
written before the upgrade of the Tevatron.

The best indirect confirmation of its presence will be the discovery of
an exotic extra family of quarks with baryon number $\pm 1$ 
\cite{FileviezPerez:2011pt}. In  the simplest such models 
\cite{Carone:1995pu,FileviezPerez:2010gw}, this fourth generation gets
its mass through the usual couplings to the Higgs, requiring its mass
to be at the electroweak scale and limited by large Yukawa couplings
leading to a Landau pole near the TeV scale.  But ref.\ 
\cite{FileviezPerez:2011pt} shows that this limitation can be removed
using vector-like quarks (from the point of view of the SM SU(2) gauge
symmetry) and giving mass to them through the VEV of the field
which breaks baryon number, $\tilde\phi$ in our model.  If the
coupling $\tilde g_B$ in (\ref{masses}) is sufficiently small, for
example $\tilde g_B \sim 0.007$, then $\langle\tilde\phi\rangle$ can be
at the TeV scale. 

Astrophysical constraints are rapidly closing in on light dark matter
models.  If $\sim 10$ GeV DM annihilates predominantly into $e^+ e^-$
with the standard relic density cross section, it is ruled out by its
distortions of the CMB \cite{Hutsi:2011vx, Galli:2011rz}.  The
$\mu^+\mu^-$ channel is still open since a large fraction of the muon
energy is converted to neutrinos which have no effect on the CMB.
Other channels have an intermediate effect between these two extremes
\cite{Hutsi:2011vx}.  In our model with two vector bosons, the
annihilation is primarily into $B$'s followed by decay into light
quarks, which appear to be still be allowed.  On the other hand, models that
produce {\it too many} neutrinos are constrained by SuperKamiokande limits
on $\chi\chi\to\nu\bar\nu$ from the sun \cite{Chen:2011vd}.  Even
in optimally isospin violating models such as we have considered here,
annihilation of $\sim 10$ GeV DM with the required cross section on
nucleons for CoGeNT/DAMA is ruled out for the $b\bar b$ channel and 
marginally allowed for $c\bar c$ and lighter quarks as in our vector
model.  These limits can be improved in the future using data from
IceCube/DeepCore \cite{Gao:2011bq}.  Finally, annihilations of light
DM in dwarf satellite galaxies of the Milky Way
that produce too many gamma rays in association with charged particles
have recently been severely constrained by Fermi data 
\cite{GeringerSameth:2011iw,collaboration:2011wa} (see also
\cite{Sandick:2011zs}).  Again, the $b\bar b$ channel is excluded but
annihilation into light quarks is still allowed.

After the first version of this paper was posted, ref.\ \cite{An:2011ck}
appeared, which considers a similar class of models.

{\bf Note added:} After completing this work we became aware of refs.\
\cite{Cao:2009uv,Lavalle:2010yw} showing that PAMELA antiproton constraints are in conflict
with a $B$ vector boson of mass greater than $2 m_p$, which would
favor the lower-$m_B$ parts of the contours of fig. 1(a).  In
addition, we discovered that the gauge boson mixing angle effects in
the present model cannot be ignored even when $\theta$ is small, due
to the occurrence of $1/m^2_{Z'}$ in the amplitudes, which scales as
$1/\theta$. This can be overcome by introducing an additional
contribution to $m_B$. We intend to address these issues in a forthcoming
publication.

\section*{Acknowledgment}

We thank B.\ Grinstein, C.\ Carone, R.\ Essig, G.\ Gelmini,  S.\
Koushiappas, J.\ Kumar, J.\ Lavalle, G.\ Moore, M.\ Pospelov,  I.\
Shoemaker, L.\ Strigari, A.\ Strumia, T.\ Volansky, Y.\ Zhang for
helpful correspondence or discussions.  JC thanks the Aspen Center for
Physics for stimulating interactions during the completion of this
work.  This work was supported in part by the Natural Sciences and
Engineering Research Council of Canada.

\appendix
\section{Kinematic functions for annihilation cross section}
\label{appA}

The functions of mass ratios appearing in the annihilation 
cross sections 
are as follows.  They were computed using Feyncalc 
\cite{Kublbeck:1992mt}.
\beqa
f_1(x_1,x_2) &=& {(1-x_2^2)^{3/2}\over \left(1-\frac12 x_1^2\right)^2}\,
	\Theta(1-x_2),
\nonumber\\  
f_2(x_1,x_2) &=& 
{\left(1-\frac12(x_1^2+x_2^2) +\frac{1}{16}(x_1^2-x_2^2)^2\right)^{3/2}\over 
	\left(1-\frac14(x_1^2+x_2^2)\right)^2}\,\Theta(2-x_1-x_2),
\nonumber\\
	f_3(x_1,x_2) &=& {\left(1+\frac12 x_2^2\right)(1-x_2^2)^{1/2}\over \left(1-\frac14
x_1^2\right)^2}\,\Theta(1-x_2)\nonumber
\eeqa 

\beqa
	f_4(x_1,x_2) &=&
{\left(1+x_1^2-\frac12 x_2^2 +\frac{1}{16}(x_1^2-x_2^2)^2\right)
\left(1-\frac12(x_1^2+x_2^2) +\frac{1}{16}(x_1^2-x_2^2)^2\right)^{1/2}
	\over 
	\left(1-\frac14(x_1^2+x_2^2)\right)^2}\nonumber\\
 &\times& \Theta(2-x_1-x_2)\nonumber
\eeqa


\begin{thebibliography}{99}


\bibitem{Bernabei:2008yi}
  R.~Bernabei {\it et al.}  [DAMA Collaboration],
  Eur.\ Phys.\ J.\  C {\bf 56} (2008) 333
  [arXiv:0804.2741 [astro-ph]].

\bibitem{Aalseth:2010vx}
  C.~E.~Aalseth {\it et al.}  [CoGeNT collaboration],
  Phys.\ Rev.\ Lett.\  {\bf 106}, 131301 (2011)
  [arXiv:1002.4703 [astro-ph.CO]].

\bibitem{Aalseth:2011wp}
  C.~E.~Aalseth {\it et al.},
  arXiv:1106.0650 [astro-ph.CO].


\bibitem{Savage:2010tg}
  C.~Savage, G.~Gelmini, P.~Gondolo, K.~Freese,
  Phys.\ Rev.\  {\bf D83}, 055002 (2011).
  [arXiv:1006.0972 [astro-ph.CO]].

\bibitem{McCabe:2011sr}
  C.~McCabe,
    [arXiv:1107.0741 [hep-ph]].


\bibitem{Ahmed:2009zw}
  Z.~Ahmed {\it et al.}  [The CDMS-II Collaboration],
  Science {\bf 327}, 1619 (2010)
  [arXiv:0912.3592 [astro-ph.CO]].

\bibitem{Ahmed:2010wy}
  Z.~Ahmed {\it et al.} [ CDMS-II Collaboration ],
  Phys.\ Rev.\ Lett.\  {\bf 106}, 131302 (2011).
  [arXiv:1011.2482 [astro-ph.CO]].

\bibitem{Angle:2009xb}
  J.~Angle {\it et al.}  [XENON10 Collaboration],
  Phys.\ Rev.\  D {\bf 80}, 115005 (2009)
  [arXiv:0910.3698 [astro-ph.CO]].

\bibitem{Aprile:2011hi}
  E.~Aprile {\it et al.}  [XENON100 Collaboration],
  arXiv:1104.2549 [astro-ph.CO].


\bibitem{Bozorgnia:2010xy}
  N.~Bozorgnia, G.~B.~Gelmini, P.~Gondolo,
  JCAP {\bf 1011}, 019 (2010).
  [arXiv:1006.3110 [astro-ph.CO]];
  JCAP {\bf 1011}, 028 (2010).
  [arXiv:1008.3676 [astro-ph.CO]].


\bibitem{Hooper:2010uy}
  D.~Hooper, J.~I.~Collar, J.~Hall, D.~McKinsey,
  Phys.\ Rev.\  {\bf D82}, 123509 (2010).
  [arXiv:1007.1005 [hep-ph]].




\bibitem{Chang:2010yk}
  S.~Chang, J.~Liu, A.~Pierce, N.~Weiner, I.~Yavin,
  JCAP {\bf 1008}, 018 (2010).
  [arXiv:1004.0697 [hep-ph]].


\bibitem{Feng:2011vu}
  J.~L.~Feng, J.~Kumar, D.~Marfatia, D.~Sanford,
    [arXiv:1102.4331 [hep-ph]].

\bibitem{TuckerSmith:2001hy}
  D.~Tucker-Smith, N.~Weiner,
  ``Inelastic dark matter,''
  Phys.\ Rev.\  {\bf D64}, 043502 (2001).
  [hep-ph/0101138].


\bibitem{Graham:2010ca}
  P.~W.~Graham, R.~Harnik, S.~Rajendran, P.~Saraswat,
  Phys.\ Rev.\  {\bf D82}, 063512 (2010).
  [arXiv:1004.0937 [hep-ph]].

\bibitem{Frandsen:2011ts}
  M.~T.~Frandsen, F.~Kahlhoefer, J.~March-Russell, C.~McCabe, M.~McCullough, K.~Schmidt-Hoberg,
    [arXiv:1105.3734 [hep-ph]].

\bibitem{Schwetz:2011xm}
  T.~Schwetz, J.~Zupan,
    [arXiv:1106.6241 [hep-ph]].


\bibitem{Fox:2011px}
  P.~J.~Fox, J.~Kopp, M.~Lisanti, N.~Weiner,
    [arXiv:1107.0717 [hep-ph]].

\bibitem{Farina:2011pw}
  M.~Farina, D.~Pappadopulo, A.~Strumia, T.~Volansky,
    [arXiv:1107.0715 [hep-ph]].





\bibitem{Fitzpatrick:2010em}
  A.~L.~Fitzpatrick, D.~Hooper, K.~M.~Zurek,
  Phys.\ Rev.\  {\bf D81}, 115005 (2010).
  [arXiv:1003.0014 [hep-ph]].


\bibitem{Essig:2010ye}
  R.~Essig, J.~Kaplan, P.~Schuster, N.~Toro,
  [arXiv:1004.0691 [hep-ph]].


\bibitem{Foot:2010rj}
  R.~Foot,
  Phys.\ Lett.\  {\bf B692}, 65-69 (2010).
  [arXiv:1004.1424 [hep-ph]].


\bibitem{Barger:2010yn}
  V.~Barger, M.~McCaskey, G.~Shaughnessy,
  Phys.\ Rev.\  {\bf D82}, 035019 (2010).
  [arXiv:1005.3328 [hep-ph]].


\bibitem{Bae:2010hr}
  K.~J.~Bae, H.~D.~Kim, S.~Shin,
  Phys.\ Rev.\  {\bf D82}, 115014 (2010).
  [arXiv:1005.5131 [hep-ph]].

\bibitem{Mambrini:2010dq}
  Y.~Mambrini,
  JCAP {\bf 1009}, 022 (2010)
  [arXiv:1006.3318 [hep-ph]].

\bibitem{Cline:2010kv}
  J.~M.~Cline, A.~R.~Frey, F.~Chen,
  Phys.\ Rev.\  {\bf D83}, 083511 (2011).
  [arXiv:1008.1784 [hep-ph]].

\bibitem{Gunion:2010dy}
  J.~F.~Gunion, A.~V.~Belikov, D.~Hooper,
    [arXiv:1009.2555 [hep-ph]].

\bibitem{Buckley:2010ve}
  M.~R.~Buckley, D.~Hooper, T.~M.~P.~Tait,
    [arXiv:1011.1499 [hep-ph]].




\bibitem{Gao:2011ka}
  X.~Gao, Z.~Kang, T.~Li,
    [arXiv:1107.3529 [hep-ph]].

\bibitem{Gondolo:2011eq}
  P.~Gondolo, P.~Ko, Y.~Omura,
    [arXiv:1106.0885 [hep-ph]].

\bibitem{DelNobile:2011je}
  E.~Del Nobile, C.~Kouvaris, F.~Sannino,
    [arXiv:1105.5431 [hep-ph]].

\bibitem{Frandsen:2011cg}
  M.~T.~Frandsen, F.~Kahlhoefer, S.~Sarkar and K.~Schmidt-Hoberg,
  arXiv:1107.2118 [hep-ph].



\bibitem{Ellis:2000ds}
  J.~R.~Ellis, A.~Ferstl, K.~A.~Olive,
  Phys.\ Lett.\  {\bf B481}, 304-314 (2000).
  [hep-ph/0001005].

\bibitem{Giedt:2009mr}
  J.~Giedt, A.~W.~Thomas and R.~D.~Young,
  Phys.\ Rev.\ Lett.\  {\bf 103}, 201802 (2009)
  [arXiv:0907.4177 [hep-ph]].

\bibitem{Williams:2011qb}
  M.~Williams, C.~P.~Burgess, A.~Maharana, F.~Quevedo,
    [arXiv:1103.4556 [hep-ph]].


\bibitem{Carone:1994aa}
  C.~D.~Carone, H.~Murayama,
  Phys.\ Rev.\ Lett.\  {\bf 74}, 3122-3125 (1995).
  [hep-ph/9411256];

\bibitem{Carone:1995pu}
  C.~D.~Carone, H.~Murayama,
  Phys.\ Rev.\  {\bf D52}, 484-493 (1995).
  [hep-ph/9501220].

\bibitem{FileviezPerez:2010gw}
  P.~Fileviez Perez, M.~B.~Wise,
  Phys.\ Rev.\  {\bf D82}, 011901 (2010).
  [arXiv:1002.1754 [hep-ph]].

\bibitem{FileviezPerez:2011pt}
  P.~Fileviez Perez, M.~B.~Wise,
  JHEP {\bf 1108}, 068 (2011).
  [arXiv:1106.0343 [hep-ph]].

\bibitem{Dulaney:2010dj}
  T.~R.~Dulaney, P.~Fileviez Perez, M.~B.~Wise,
  Phys.\ Rev.\  {\bf D83}, 023520 (2011).
  [arXiv:1005.0617 [hep-ph]].

\bibitem{Pospelov:2011ha}
  M.~Pospelov,
    [arXiv:1103.3261 [hep-ph]].

\bibitem{Kaplan:2009ag}
  D.~E.~Kaplan, M.~A.~Luty, K.~M.~Zurek,
  Phys.\ Rev.\  {\bf D79}, 115016 (2009).
  [arXiv:0901.4117 [hep-ph]].


\bibitem{O'Connell:2006wi}
  D.~O'Connell, M.~J.~Ramsey-Musolf, M.~B.~Wise,
  Phys.\ Rev.\  {\bf D75}, 037701 (2007).
  [hep-ph/0611014].


\bibitem{Chen:2009ab}
  F.~Chen, J.~M.~Cline, A.~R.~Frey,
  Phys.\ Rev.\  {\bf D80}, 083516 (2009).
  [arXiv:0907.4746 [hep-ph]].


\bibitem{Cheung:2009qd}
  C.~Cheung, J.~T.~Ruderman, L.~-T.~Wang, I.~Yavin,
  Phys.\ Rev.\  {\bf D80}, 035008 (2009).
  [arXiv:0902.3246 [hep-ph]].

\bibitem{Bjorken:2009mm}
  J.~D.~Bjorken, R.~Essig, P.~Schuster, N.~Toro,
  Phys.\ Rev.\  {\bf D80}, 075018 (2009).
  [arXiv:0906.0580 [hep-ph]].

\bibitem{Essig:2009nc}
  R.~Essig, P.~Schuster, N.~Toro,
  Phys.\ Rev.\  {\bf D80}, 015003 (2009).
  [arXiv:0903.3941 [hep-ph]].


\bibitem{Essig:2010xa}
  R.~Essig, P.~Schuster, N.~Toro, B.~Wojtsekhowski,
  JHEP {\bf 1102}, 009 (2011).
  [arXiv:1001.2557 [hep-ph]].

\bibitem{Merkel:2011ze}
  H.~Merkel {\it et al.} [ A1 Collaboration ],
  Phys.\ Rev.\ Lett.\  {\bf 106}, 251802 (2011).
  [arXiv:1101.4091 [nucl-ex]].

\bibitem{Archilli:2011nh}
  F.~Archilli, D.~Babusci, D.~Badoni, I.~Balwierz, G.~Bencivenni, C.~Bini, C.~Bloise, V.~Bocci {\it et al.},
    [arXiv:1107.2531 [hep-ex]].

\bibitem{Aranda:1998fr}
  A.~Aranda, C.~D.~Carone,
  Phys.\ Lett.\  {\bf B443}, 352-358 (1998).
  [hep-ph/9809522].

\bibitem{Graesser:2011vj}
  M.~L.~Graesser, I.~M.~Shoemaker, L.~Vecchi,
    [arXiv:1107.2666 [hep-ph]].

  
\bibitem{Hutsi:2011vx}
  G.~Hutsi, J.~Chluba, A.~Hektor, M.~Raidal,
    [arXiv:1103.2766 [astro-ph.CO]].

\bibitem{Galli:2011rz}
  S.~Galli, F.~Iocco, G.~Bertone, A.~Melchiorri,
  Phys.\ Rev.\  {\bf D84}, 027302 (2011).
  [arXiv:1106.1528 [astro-ph.CO]].

\bibitem{Chen:2011vd}
  S.~-L.~Chen, Y.~Zhang,
    [arXiv:1106.4044 [hep-ph]].

\bibitem{Gao:2011bq}
  Y.~Gao, J.~Kumar, D.~Marfatia,
    [arXiv:1108.0518 [hep-ph]].

\bibitem{GeringerSameth:2011iw}
  A.~Geringer-Sameth, S.~M.~Koushiappas,
    [arXiv:1108.2914 [astro-ph.CO]].

\bibitem{collaboration:2011wa}
  Fermi/LAT ollaboration,
    [arXiv:1108.3546 [astro-ph.HE]].

\bibitem{Sandick:2011zs}
  P.~Sandick, J.~Diemand, K.~Freese and D.~Spolyar,
  arXiv:1108.3820 [astro-ph.CO].

\bibitem{An:2011ck}
  H.~An and F.~Gao,
  arXiv:1108.3943 [hep-ph].

\bibitem{Cao:2009uv}
  Q.~-H.~Cao, I.~Low, G.~Shaughnessy,
  Phys.\ Lett.\  {\bf B691}, 73-76 (2010).
  [arXiv:0912.4510 [hep-ph]].

\bibitem{Lavalle:2010yw}
  J.~Lavalle,
  Phys.\ Rev.\  D {\bf 82}, 081302 (2010)
  [arXiv:1007.5253 [astro-ph.HE]].

\bibitem{Kublbeck:1992mt}
  J.~Kublbeck, H.~Eck, R.~Mertig,
  Nucl.\ Phys.\ Proc.\ Suppl.\  {\bf 29A}, 204-208 (1992).


\end{thebibliography}
\end{document}